\newif{\ifjournal}
  \journalfalse

\ifjournal
  \documentclass{aa}
  \usepackage{graphicx,times,amssymb,natbib}
  \renewcommand{\d}{\mathrm{d}}
  \authorrunning{Bartelmann \& Meneghetti}
  \titlerunning{Do arcs require flat halo cusps?}
\else
  \documentclass{paper}
\fi

\begin{document}

\title{Do arcs require flat halo cusps?}
\ifjournal
  \author{Matthias Bartelmann\inst{1}, Massimo Meneghetti\inst{2}}
  \institute
    {$^1$ ITA, Universit\"at Heidelberg, Tiergartenstr.~15, D--69121
     Heidelberg, Germany\\
     $^2$ Dipartimento di Astronomia, Universit\`a di Padova, vicolo
     dell'Osservatorio 5, I--35122 Padova, Italy}
\else
  \author{Matthias Bartelmann$^1$, Massimo Meneghetti$^2$\\
    $^1$ ITA, Universit\"at Heidelberg, Tiergartenstr.~15, D--69121
    Heidelberg, Germany\\
    $^2$ Dipartimento di Astronomia, Universit\`a di Padova, vicolo
    dell'Osservatorio 5, I--35122 Padova, Italy}
\fi

\date{\today}

\newcommand{\abstext}
  {It was recently claimed that several galaxy clusters containing
   radial and tangential gravitational arcs and having a measured
   velocity-dispersion profile for the brightest cluster galaxy had to
   have central density profiles considerably flatter than those found
   in CDM cluster simulations. Using a simple analytic mass model, we
   confirm this result \emph{for axially symmetric} mass
   distributions, but show that steep density profiles are well in
   agreement with the cluster requiring the flattest axially symmetric
   profile once even small deviations from axial symmetry are
   introduced.}

\ifjournal
  \abstract{\abstext}
\else
  \begin{abstract}
    \abstext
  \end{abstract}
\fi

\maketitle

\section{Introduction}

Are observed gravitational arcs in galaxy clusters compatible with the
density profiles produced in CDM simulations, which consistently find
that the dark-matter density increases towards halo centres as
$r^{-1}$ or steeper? Building upon a suggestion by \cite{MI95.1},
\citeauthor{SA03.1} (\citeyear{SA03.1}, hereafter STSE) have recently
analysed six galaxy clusters containing tangential arcs, three of
which also contain radial arcs. Apart from the lensing data, the
method uses constraints on the mass profile derived from the dynamics
of the central cluster galaxies, specifically from its velocity
dispersion profile (see also \citealt{SA02.1}).

The method sets strong constraints. In a cluster showing both radial
and tangential arcs, the velocity dispersion measurement essentially
fixes the mass divided by the radius. Radial arcs constrain the slope
of the projected mass profile at their location, and tangential arcs
constrain the total mass enclosed by their radial distance from the
cluster centre.

Using this technique, STSE find that their sample of six clusters is
incompatible with dark-matter density profiles proportional to
$r^{-1}$ or steeper, but consistently require profiles as flat as
$r^{-0.5}$.

If true, this result would be of great importance for CDM because such
flat profiles are not found in simulations. However, the analysis by
STSE is based on the assumption of axial symmetry. We re-analyse their
constraints here for the cluster Abell~383, which is the one
apparently requiring the most significant discrepancy between
simulated CDM density profiles. In doing so, we describe the lens
model and its parameters in Sect.~2 and illustrate the basic reason
why the method strongly prefers a shallow central density
profile. Next, we introduce ellipticity in Sect.~3 and illustrate its
substantial impact. We conclude in Sect.~4 that even moderate
ellipticity can easily remove the discrepancy between the lensing
observations in Abell~383 and the typical CDM halo profiles.

Like STSE, we use a $\Lambda$CDM cosmological model with matter
density $\Omega_0=0.3$, cosmological constant $\Omega_\Lambda=0.7$,
and Hubble constant $H_0=65\,\mathrm{km\,s^{-1}\,Mpc^{-1}}$.

\section{Lens Model}

Our lens model is adapted from STSE. It consists of a model for the
dark-matter halo,
\begin{equation}
  \rho_\mathrm{h}(x)=\frac{\rho_\mathrm{s}}{x^\beta(1+x)^{3-\beta}}\;,
\label{eq:1}
\end{equation}
which is the profile found by \cite{NA97.1} for $\beta=1$. Steeper
profiles with $\beta\to1.5$ are found by several other groups, among
them \cite{MO98.1,JI00.2,KL01.1}. The radial coordinate
$x=r/r_\mathrm{s}$ is the physical radius divided by a scale radius
$r_\mathrm{s}$, which STSE assume to be
$r_\mathrm{s}=400\,\mathrm{kpc}$. We adopt this value because the
results are insensitive to it.

In addition, the brightest cluster galaxy, assumed to be concentric
with the cluster, adds matter density to the cluster centre. Following
the light profile, it is assumed to have a \cite{JA83.1} density
profile,
\begin{equation}
  \rho_\mathrm{g}(\bar{x})=
  \frac{\rho_\mathrm{J}}{\bar{x}^2(1+\bar{x})^2}\;,
\label{eq:2}
\end{equation}
where now $\bar{x}=r/r_\mathrm{J}$ is the radius in units of the Jaffe
radius $r_\mathrm{J}$. Fitting to the light profile of the brightest
cluster galaxy, STSE find $r_\mathrm{J}=r_\mathrm{e}/0.76$, where
$r_\mathrm{e}$ is the usual effective radius of an $r^{1/4}$
surface-brightness profile. For Abell~383,
$r_\mathrm{e}=46.75\pm2.04\,\mathrm{kpc}$.

Lensing properties are straightforwardly derived from these density
profiles. The convergence $\kappa$ for the dark-matter halo is
\begin{equation}
  \kappa_\mathrm{h}(x)=2\,\kappa_\mathrm{s}\,x^{1-\beta}\,
  \int_0^{\pi/2}\frac{\sin\theta\d\theta}{(x+\sin\theta)^{3-\beta}}
\label{eq:3}
\end{equation}
\citep{WY01.1}; the special case for the NFW profile, $\beta=1$, was
derived in \cite{BA96.1}. The factor $\kappa_\mathrm{s}$ is defined by
\begin{equation}
  \kappa_\mathrm{s}\equiv\rho_\mathrm{s}r_\mathrm{s}\,
  \Sigma_\mathrm{cr}^{-1}\;,
\label{eq:3a}
\end{equation}
with the usual critical surface-mass density for lensing
$\Sigma_\mathrm{cr}$. For the Jaffe profile,
\begin{equation}
  \kappa_\mathrm{g}=\kappa_\mathrm{J}\,\left[
    \frac{\pi}{x}+\frac{2}{1-x^2}\,\left(
      1-\frac{2-x^2}{\sqrt{1-x^2}}\mathrm{acosh}\frac{1}{x}
    \right)
  \right]\;,
\label{eq:4}
\end{equation}
\citep{JA83.1}, with
$\kappa_\mathrm{J}\equiv\rho_\mathrm{J}r_\mathrm{J}\,\Sigma_\mathrm{cr}^{-1}$.
Deflection angles are derived from
\begin{equation}
  \alpha(x)=\frac{2}{x}\,\int_0^x\,y\kappa(y)\d y\;.
\label{eq:5}
\end{equation}
This needs to be computed numerically for the dark-matter profile,
while
\begin{equation}
  \alpha_\mathrm{J}(x)=\kappa_\mathrm{J}\,\left[
    \pi-\frac{2x\,\mathrm{acosh}(x^{-1})}{\sqrt{1-x^2}}
  \right]
\label{eq:6}
\end{equation}
for the Jaffe profile. Given the deflection-angle profile $\alpha(x)$,
the radial and tangential eigenvalues are
\begin{equation}
  \lambda_\mathrm{r}(x)=1-\frac{\d\alpha(x)}{\d x}\;,\quad
  \lambda_\mathrm{t}(x)=1-\frac{\alpha(x)}{x}\;,
\label{eq:7}
\end{equation}
respectively (see, e.g., \citealt{SC92.1,NA99.1}). Radial or
tangential critical curves are found where $\lambda_\mathrm{r}=0$ or
$\lambda_\mathrm{t}=0$. A useful relation for axially symmetric lens
models is
\begin{equation}
  \lambda_\mathrm{r}(x)=2\kappa(x)-\frac{\alpha(x)}{x}\;.
\label{eq:8}
\end{equation}

In our application, source and lens redshifts together with the
cosmological model fix the critical surface-mass density
$\Sigma_\mathrm{cr}$. The Jaffe radius $r_\mathrm{J}$ is fixed by
fitting the surface-brightness profile. The mass of the brightest
cluster galaxy is determined by the central velocity dispersion as
described below. The scale radius of the dark-matter profile is kept
fixed at $r_\mathrm{s}=400\,\mathrm{kpc}$. The only remaining
parameter is the dark-matter density scale $\rho_\mathrm{s}$, which we
express by the ratio $\mu$ between the masses contributed by the
cluster- and galaxy density profiles within the Jaffe radius. This
ratio will effectively scale the cluster mass relative to the fixed
galaxy mass.

The circular velocity for an spherically symmetric density profile is
\begin{equation}
  v_\mathrm{rot}^2=\frac{G\,M(r)}{r}\;.
\label{eq:9}
\end{equation}
For the dark-matter profile, $M(r)$ is steeper than $r$, hence
$v_\mathrm{rot}\to0$ for $r\to0$ in the dark-matter profile
only. Thus, the measured circular velocity near the centre of the
brightest cluster galaxy must be dominated by the galaxy itself. Well
within the Jaffe radius, the Jaffe profile is isothermal with
$\rho_\mathrm{g}\propto r^{-2}$. Thus, the velocity dispersion profile
is expected to be flat close to the cluster centre, and the velocity
dispersion is approximately related to the circular velocity by
\begin{equation}
  \sigma_v=\sqrt{2}\,v_\mathrm{rot}\;.
\label{eq:10}
\end{equation}
Through Eqs.~(\ref{eq:9}) and (\ref{eq:10}), the central velocity
dispersion thus fixes the mass contained in the Jaffe profile. The
assumed mass ratio between the masses of the dark halo and the galaxy
profile, and the central slope $\beta$ of the dark-matter density
profile, thus completely determine the total density profile composed
of a galaxy- and a dark-matter profile.

Specifically, aiming at the cluster Abell~383 which produced the most
significant deviation from the numerically simulated dark-matter
profiles in the study by STSE, we adopt the fixed parameters listed in
Tab.~\ref{tab:1}.

\begin{table}
\caption{Fixed parameters assumed for the lensing analysis in this
  paper, taken from STSE. The two remaining parameters, i.e.~the
  density profile slope $\beta$ and the ratio of masses $\mu$
  contributed by the dark-matter and galaxy density profiles contained
  within the Jaffe radius, are taken as free parameters.}
\label{tab:1}
\begin{center}
\begin{tabular}{|l|l|r|}
  \hline
  lens redshift & $z_\mathrm{d}$ & $0.189$ \\
  source redshift & $z_\mathrm{s}$ & $1.01$ \\
  effective radius & $r_\mathrm{e}$ & $46.75\,\mathrm{kpc}$ \\
  velocity dispersion &
    $\sigma_\mathrm{v}$ & $250\,\mathrm{km\,s^{-1}}$ \\
  \hline
\end{tabular}
\end{center}
\end{table}

\begin{figure}[ht]
  \includegraphics[width=\hsize]{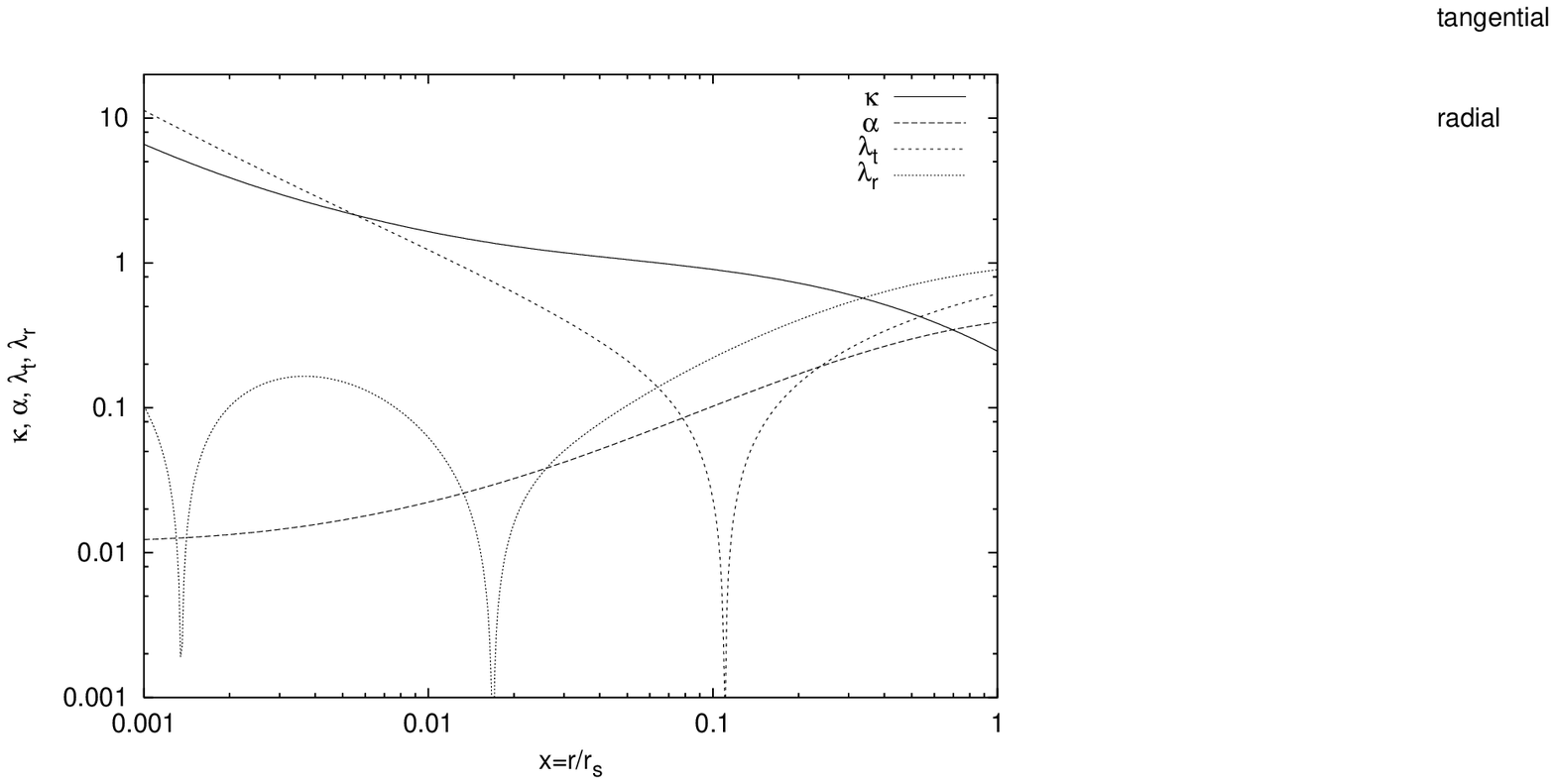}
\caption{Radial profiles are shown for the lensing convergence
  $\kappa$, the deflection angle $\alpha$, and the tangential and
  radial eigenvalues of the lens mapping $\lambda_\mathrm{t}$ and
  $\lambda_\mathrm{r}$, respectively. The lens model is axially
  symmetric and composed of a central Jaffe profile and a dark matter
  profile with a central slope of $-\beta=-0.5$.}
\label{fig:1}
\end{figure}

The convergence, deflection-angle and radial and tangential eigenvalue
profiles $\kappa(x)$, $\alpha(x)$, $|\lambda_\mathrm{r}(x)|$ and
$|\lambda_\mathrm{t}(x)|$ are displayed in Fig.~\ref{fig:1} for
$\beta=0.5$ and mass ratio $\mu=60$. The figure illustrates that the
(projected) density profile near the cluster centre is dominated by
the near-isothermal slope of the brightest cluster galaxy. Near the
Jaffe radius, it flattens towards the central slope of the dark-matter
density profile, and then steepens towards $\kappa\propto x^{-2}$ as
$x$ approaches unity. Interestingly, the radial eigenvalue profile has
two roots, indicating the presence of two radial critical curves. For
Abell~383, the inner radial critical curve is too close to the cluster
centre for having any practical relevance, but clusters at more
favourable redshifts might show signatures of a double radial critical
curve. This is an interesting feature of the combination of a steep,
near-isothermal galaxy profile embedded into a relatively flat
dark-matter halo.

Having defined the axially symmetric lens model, the constraints
imposed by the central velocity dispersion, the radial and the
tangential arcs are straightforwardly understood. As mentioned before,
the central velocity dispersion is almost exclusively contributed by
the mass associated with the brightest cluster galaxy because $M/r$
tends to zero for $r\to0$ for the flatter dark-matter density
profile. In contrast, the tangential arc is located at a radius which
encloses a mean surface-mass density of unity. A first constraint thus
derives from the requirement to have the central cluster mass
dominated by the brightest cluster galaxy, and yet to have sufficient
mass in the dark-matter halo to produce tangential arcs at relatively
large cluster-centric radii. In the axially symmetric models, this is
achieved by flattening the dark-matter density profile.

A second constraint is imposed by the radial critical curve, where the
derivative of the deflection angle reaches unity. For the Jaffe
profile alone, the deflection angle is flat. It steepens as the
total density profile becomes flatter at radii where the dark matter
starts dominating, then flattens again as the cluster-centric radius
approaches the scale radius of the dark-matter profile. If the
dark-matter density profile is relatively flat, the increase of the
deflection-angle slope occurs closer to the brightest cluster galaxy
than for a steeper dark-matter profile. The second constraint thus
derives from the requirement of having a radial arc rather close to
the brightest cluster galaxy, while the first constraint requires a
tangential arc rather far away from the cluster centre.

\begin{figure}[ht]
  \includegraphics[width=\hsize]{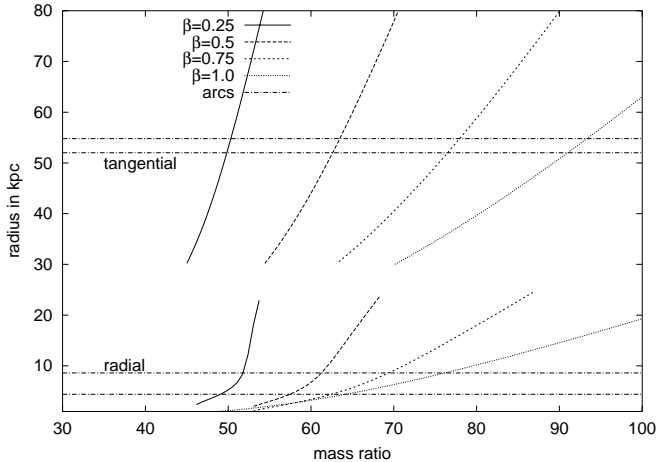}
\caption{Illustration of the radial and tangential critical radii for
  an axially symmetric lens model composed of a central Jaffe profile
  and a dark matter profile with four different values for the central
  slope, $\beta=\{0.25,0.5,0.75,1.0\}$, respectively, as indicated in
  the plot. The abscissa shows the mass ratio between the cluster
  component and the galaxy component within the Jaffe radius, the
  ordinate is the cluster-centric radius in kpc. The two horizontal
  stripes mark the locations of the radial and tangential arcs, as
  labelled. The curves in the upper and lower halves of the plot mark
  the tangential and radial critical radii, respectively.}
\label{fig:2}
\end{figure}

Figure~\ref{fig:2} illustrates the situation. The abscissa is the mass
ratio between dark and luminous constituents within the Jaffe
radius. The ordinate is the cluster-centric radius in kpc. The
horizontal bars mark the radial and tangential arc locations in
Abell~383 with their respective uncertainties. The curves in the upper
and lower halves of the figure show the tangential and radial critical
radii, respectively, for dark-matter profiles with four different
central slopes, $\beta\in\{0.25,0.5,0.75,1.0\}$, as marked in the
plot.

A model can explain both the radial and the tangential arc if there is
a single mass ratio for which the radial critical curve falls into the
lower horizontal band, and the tangential critical curve falls into
the upper horizontal band. Figure~\ref{fig:2} confirms the result by
STSE regarding Abell~383. Only for shallow central density profiles,
$\beta\sim0.5$, can the position of both the tangential and the radial
arc be understood. For a mass ratio near 60, the lens model has radial
and tangential critical curves in the observed ranges. Steeper mass
profiles, e.g.~the NFW model with $\beta=1$, either have the
tangential arc in the right range if the mass ratio is $\sim90$, but
then the radial arc is too distant from the cluster centre, or the
radial arc location is reproduced if the mass ratio is $\sim70$, but
then the tangential arc is way too close to the cluster centre. The
core of the problem is thus that the location of the radial arc
requires a relatively low cluster mass, and then a steep mass profile
forces the tangential arc too close to the cluster centre.

\section{Ellipticity}

\subsection{Model and numerical results}

These conclusions are valid for axially symmetric lens models. As we
shall show now, the situation changes considerably if deviations from
axial symmetry are allowed.

As a simple model for asymmetry, we deform the lens model such that
iso-contour lines of the lensing potential $\psi$ are ellipses. We thus
introduce the radial coordinate
\begin{equation}
  \bar{x}=\left[
    (1-\epsilon)\,x_1^2+\frac{x_2^2}{1-\epsilon}
  \right]^{1/2}
\label{eq:11}
\end{equation}
and replace $\psi(x)$ by $\psi(\bar{x})$. Being the gradient of
$\psi$, the deflection angle $\vec\alpha$ now has the components
\begin{equation}
  \alpha_1(x_1,x_2)=\frac{\alpha(\bar{x})\,x_1}{\bar{x}}(1-\epsilon)
  \;,\quad
  \alpha_2(x_1,x_2)=\frac{\alpha(\bar{x})\,x_2}{(1-\epsilon)\bar{x}}\;,
\label{eq:12}
\end{equation}
where $\alpha(\bar{x})$ is the deflection-angle profile of the axially
symmetric lens taken at $\bar{x}$. The axially symmetric case is
recovered for $\epsilon=0$.

Elliptical distortions of the lensing potential lead to
dumbbell-shaped surface-mass distributions if $\epsilon$ becomes
large, $\epsilon\gtrsim0.2$ day, depending on the density
profile. This is certainly unwanted for galaxy-sized lenses, but not
necessarily for cluster lenses which are often highly structured. In
any case, we shall see below that the impact of a small ellipticity
$\epsilon\ll1$ on the location of the critical curves is identical for
lenses with elliptical iso-potential curves and axially symmetric
lenses embedded into external shear.

\begin{figure}[ht]
  \includegraphics[width=\hsize]{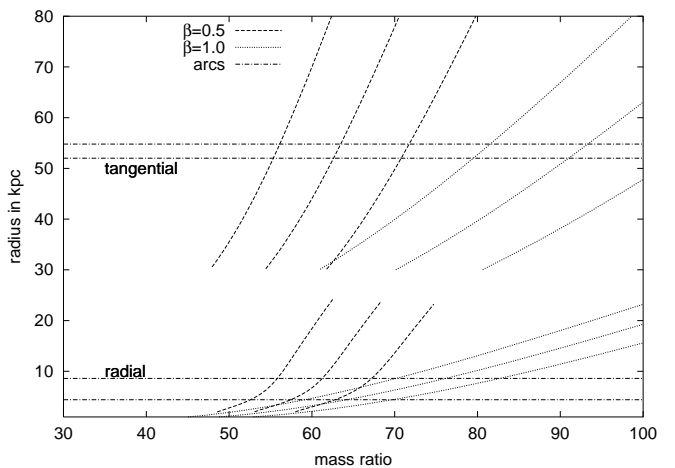}
\caption{Changes in the location of the radial and tangential critical
  curves in response to a moderate elliptical distortion of the
  lensing potential. The figure is arranged to resemble
  Fig.~\ref{fig:2}, but showing results for $\beta=0.5$ and
  $\beta=1.0$ only for clarity. Three curves are shown for each
  profile slope. The central curve shows the critical radius for axial
  symmetry, the other two curves indicate the range of radii of the
  critical curves for an ellipticity parameter of $\epsilon=0.1$. Even
  moderate ellipticity widens the radial ranges such that the observed
  radial and tangential critical radii can quite easily be reached
  even for steep profiles.}
\label{fig:3}
\end{figure}

Ellipticity stretches the tangential critical curves along the major
axis of the ellipse and shrinks it along the minor axis, and deforms
the radial critical curve in the perpendicular direction. This implies
that the cluster-centric distance of the critical curves now covers a
range of radii. This range is surprisingly wide even for small
ellipticities, as Fig.~\ref{fig:3} illustrates.

The figure is arranged in the same way as Fig.~\ref{fig:2} and also
specialised for the cluster Abell~383. The radius from the cluster
centre is plotted against the mass ratio between cluster and brightest
cluster galaxy for different central slopes $\beta$ of the cluster
density profile. The observed locations of the tangential and radial
arcs are marked as horizontal bars. For clarity, we now show results
for two values of $\beta$ only, $\beta\in\{0.5,1\}$, but illustrate
with three curves for each $\beta$ the radial range covered by the
tangential and radial critical curves for a small ellipticity
parameter $\epsilon=0.1$. For a fixed mass ratio, the radial range is
given by the vertical distance between the top and bottom curves of
the same type. Conversely, for a fixed radius, the horizontal distance
between the left and right curves of the same type show the range of
mass ratios for which arcs at that radius can be produced somewhere
along the respective critical curves. The central curves reproduce the
critical radii for the axially symmetric case.

Evidently, the impact of the small ellipticity $\epsilon$ on the
location of the critical curves is quite large. For the shallow
profile, $\beta=0.5$, and a mass ratio of $60$, the tangential
critical radius ranges from $27$ to $70$~kpc, while the axially
symmetric result is $45$. For the same mass ratio and density-profile
slope of $\beta=0.5$, the radial critical radius ranges between $3$
and $19$~kpc around the axially-symmetric value of $7$~kpc. Thus, even
the low ellipticity of $\epsilon=0.1$ makes the location of the
tangential and radial critical curves vary by about a factor of $1.5$
and more than a factor of two, respectively, around the
cluster. Interestingly, the radial ranges for tangential and radial
critical curves now overlap even for the steep profile with
$\beta=1$. For a mass ratio of $\sim80$, the observed positions of
both the radial and the tangential arcs fall within the ranges allowed
by the model.

\subsection{Analytic description\label{sec:3.2}}

\begin{figure}[ht]
  \includegraphics[width=\hsize]{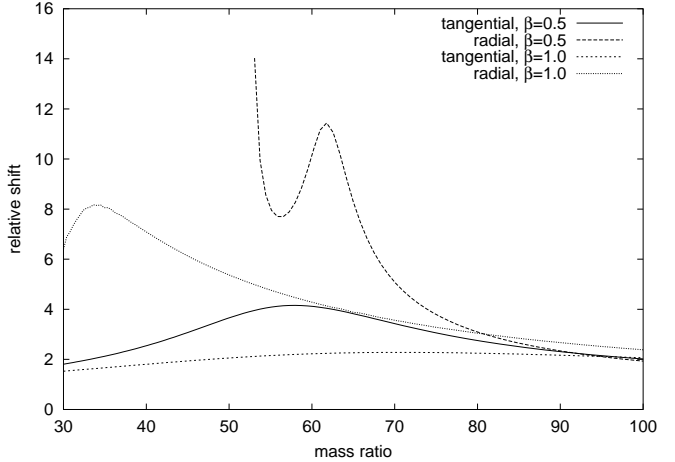}
\caption{Relative shift per unit ellipticity, $\delta x/(\epsilon x)$,
  of the radial and tangential critical curves caused by external
  shear or internal ellipticity. Results for two values of the central
  density-profile slope $\beta$ are shown. For example, for a mass
  ratio of $60$, $\beta=0.5$ and $\epsilon=0.1$, the radial critical
  curve shifts by $\sim100\%$, the tangential curve by $\sim40\%$.}
\label{fig:4}
\end{figure}

These results can be reproduced analytically. For coordinate axes
aligned with the elliptical iso-potential contours, the minimum and
maximum values of the critical radii occur on the axes. The
intersection of the tangential critical curve with the $x_1$-axis
satisfies
\begin{equation}
  1-\frac{\alpha(\bar{x})}{(1-\epsilon)\bar{x}}=0\;.
\label{eq:13}
\end{equation}
Expanding around the solution for the critical radius in the axially
symmetric case, which satisfies
\begin{equation}
  1-\frac{\alpha(x)}{x}=0\;,
\label{eq:13a}
\end{equation}
and assuming $\epsilon\ll1$, we find the amount $\delta x$ by which
the tangential critical curve is shifted on the $x_1$-axis relative to
the axially symmetric critical radius. Repeating the calculation for
the $x_2$ axis, we obtain
\begin{equation}
  \delta x_\mathrm{t}=\pm\frac{\epsilon\,x}{2\,(\kappa_\mathrm{t}-1)}
\label{eq:14}
\end{equation}
where the $+$ and $-$ signs apply to the $x_1$ and $x_2$ axes,
respectively, and $\kappa_\mathrm{t}$ is the convergence at the
tangential critical radius of the axially symmetric model. Repeating
this analysis for the radial critical curve yields
\begin{equation}
  \delta x_\mathrm{r}=\pm\frac{\epsilon\,x}
    {2\,(\kappa_\mathrm{r}-1+\kappa^\prime_\mathrm{r}x)}\;,
\label{eq:15}
\end{equation}
where $\kappa_\mathrm{r}$ and $\kappa^\prime_\mathrm{r}$ are the
convergence and its radial derivative at the radial critical radius of
the unperturbed lens model. Figure~\ref{fig:4} shows the relative
shifts for unit ellipticity, i.e.~$\delta x_\mathrm{t,r}/(\epsilon
x)$, for the two central slopes $\beta\in\{0.5,1\}$.

The figure shows that, for a mass ratio of $60$ and a central density
slope of $\beta=0.5$, the relative shifts for $\epsilon=0.1$ are of
order $100\%$ for the radial critical curve, and $40\%$ for the
tangential critical curve, confirming the numerical results
illustrated in Fig.~\ref{fig:3}. For the steeper profile with
$\beta=1$, the respective relative changes are of order $40\%$ and
$20\%$ for the same mass ratio and ellipticity.

It is straightforward to show that the first-order results
(\ref{eq:14}) and (\ref{eq:15}) remain valid if the lens model is not
itself deformed, but embedded into an external shear $\gamma$. In that
case, $\gamma$ simply replaces $\epsilon$ in these equations which
are otherwise unchanged.

Equations~(\ref{eq:14}) and (\ref{eq:15}) show that the amount by
which ellipticity or shear shift the critical curves depends highly
sensitively on the slope of the convergence profile $\kappa(x)$. For a
singular isothermal sphere, for instance, $\kappa=0.5$ at the location
of the tangential critical curve, which is at $x=1$. Thus,
$\delta_\mathrm{t}=\pm\epsilon$ in this case. Flatter profiles,
however, have $\kappa$ closer to unity in Eq.~(\ref{eq:14}) because
the tangential critical radius encloses a mean convergence of
unity. The relative shift of the tangential critical curve is thus
amplified for flatter density profiles, as illustrated in
Fig.~\ref{fig:4} for the two choices of $\beta$. As the figure also
shows, the situation is more extreme for the radial critical
curve. Flatter profiles thus much more sensitive to external shear or
internal ellipticity than steeper profiles.

\subsection{Numerical Example}

We illustrate the analytic results obtained above using a numerically
simulated galaxy cluster as a lens. It is located at redshift $z=0.3$
and has a mass of $M=5\times10^14\,h^{-1}\,M_\odot$. The cluster was
taken from a large-scale numerical simulation of the $\Lambda$CDM
model with parameters $\Omega_0=0.3$, $\Omega_\Lambda=0.7$,
$H_0=70\,\mathrm{km\,s^{-1}\,Mpc^{-1}}$ and normalisation
$\sigma_8=0.9$. It is one of the clusters produced by the GIF
collaboration \citep{KA99.2} used in earlier related studies
\citep{BA98.2,ME00.1,ME03.1,ME03.2}. The particle mass in the
simulation is $1.4\times10^{10}\,h^{-1}M_\odot$ and the gravitational
softening length was set to $30\,h^{-1}\,\mbox{kpc}$.

We first use the cluster as it is, i.e.~with the asymmetry and
substructure produced by the simulation, and then progressively smooth
and circularise it by computing its azimuthally-averaged density
profile, subtracting it from the cluster, smoothing the residual
density distribution by a varying amount, and finally adding the
axially symmetric density profile back on the smoothed
residuals. Using a normalised smoothing kernel, this procedure
preserves the total mass and the mean density profile of the cluster.

\begin{figure}[ht]
  \includegraphics[width=\hsize]{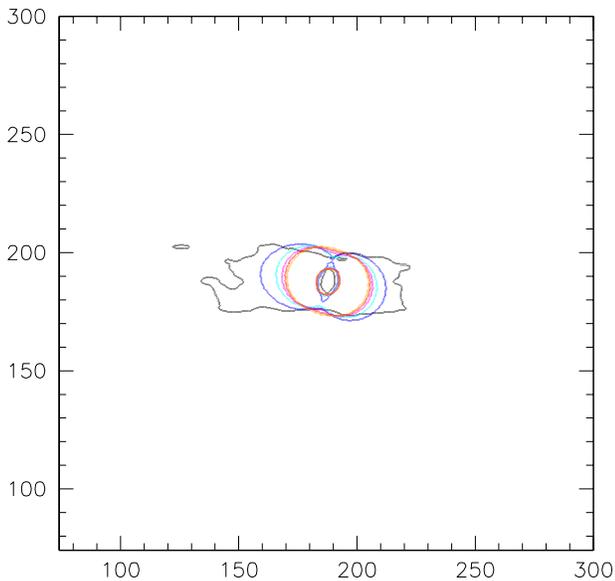}
\caption{Critical curves of a simulated cluster in various stages of
  smoothing. As described in the text, the smoothing procedure
  conserves the total cluster mass and its density profile by
  construction. The ragged line is the original critical curve. Even
  moderate smoothing makes the critical curves shrink and considerably
  reduces the radial range where tangential and radial arcs can be
  found.}
\label{fig:5}
\end{figure}

Figure~\ref{fig:5} illustrates the effect of the progressive smoothing
on the critical curves. The ragged lines show the critical curve of
the original clusters, while the approximately elliptical, smooth
curves are the critical curves of the cluster after
smoothing. Evidently, smoothing makes the critical curves shrink
considerably, and, more importantly for our discussion, the radial
range covered by the critical curves narrows substantially while the
mean density profile remains entirely unchanged by
construction. Arguments based on the cluster-centric distance of
radial and tangential arcs thus need to take the detailed cluster
structure into account.

\section{Summary}

Galaxy clusters containing radial and tangential gravitational arcs
and a brightest cluster galaxy with a measured velocity dispersion
apparently require significantly flatter density profiles than
obtained in numerical simulations of CDM cosmologies. As shown here,
this is essentially caused by the large observed cluster-centric
distances of tangential arcs, which require fairly flat density
profiles given the central constraints of radial arcs and the
velocity-dispersion measurement. Using a simple analytic mass model,
we can confirm the results by STSE, \emph{provided the lensing mass
distribution is axially symmetric}.

Allowing deviations from axial symmetry, the results radically
change. We have chosen to introduce asymmetry by elliptically
distorting the lensing potential, but showed that embedding the
axially symmetric lens into external shear has identical consequences
in the limit of small ellipticity or shear. Based on these results, we
have shown that the particular cluster which most significantly
required a flat density profile in the analysis by STSE, Abell~383, is
well compatible with an NFW profile ($\beta=1$) even for the small
ellipticity of $\epsilon=0.1$.

Critical curves caused by flat density profiles are extremely
sensitive to distortions, as demonstrated in Sect.~\ref{sec:3.2} and
illustrated in Fig.~\ref{fig:4}. Shifting tangential critical curves
by 20\% to 40\%, and radial critical curves by 50\% to 100\% even with
an ellipticity or shear of only $0.1$ is possible in particular for
the profiles as flat as advertised by STSE. This is also the reason
why the analysis of cluster ellipticity carried out by STSE themselves
concluded that ellipticity had a negligible effect on their results:
their lens model used components with isothermal density profiles
which are much less sensitive to external shear or distortions, as
illustrated by Eqs.~(\ref{eq:14}) and (\ref{eq:15}).

We conclude that radial and tangential arcs in clusters do not rule
out central density profiles as steep as found in CDM simulations once
effects of asymmetry and shear are taken into account. We will extend
our analysis towards numerically simulated clusters in a forthcoming
paper.

\bibliography{../../TeXMacro/master}
\bibliographystyle{../../TeXMacro/aa}

\end{document}